\documentclass[prl,twocolumn]{revtex4}
\usepackage{amssymb}
\usepackage{amsmath}
\usepackage{epsfig}
\usepackage{bm}

\begin{document}

\title{ Giant Hall effect in the ballistic transport of two-dimensional electrons }
\author{ Yu. O. Alekseev $^1$ and A. P. Dmitriev $^2$ }
\affiliation{$^1$ Lyceum "Physical-Technical High School", Hlopina 8-3A, ~St. Petersburg, 194021, Russia\\
	$^2$Ioffe  Institute, Politekhnicheskaya 26, 194021,   St.~Petersburg,   Russia}

 \begin{abstract}

	We have studied magnetotransport of a degenerate two-dimensional electron gas in a Hall sample in the Knudsen regime, when the mean free paths of electrons with respect to their collisions with each other and with impurities are much larger than the width of the sample. In contrast to the usually considered symmetric sample, whose both its edges reflect electrons diffusely, we considered an asymmetric sample, one edge of which reflects them diffusely, while the other specularly. It is shown that in such structure in low magnetic fields the Hall coefficient is parametrically large in comparison with its standard value. Also the situation is discussed when all types of scattering can be neglected except for scattering at the edges of the sample.
		

 \end{abstract}

 \maketitle

\section{Introduction}
In recent years, in connection with the impressive progress in the creation of two-dimensional systems with a record mobility of carriers, interest in the theoretical study of the effect of interparticle interaction on transport phenomena has sharply increased [1-11], the role of which in dirty systems with low mobility is insignificant. At the same time, research is being conducted in two directions. On the one hand, the hydrodynamic regime of electron transport is being intensively studied, which is realized, apparently, in experiments on giant temperature-dependent magnetoresistance in ultrapure semiconductor and graphene samples [12-24]. On the other hand, ballistic and intermediate between ballistic and hydrodynamic regimes are eximined extensively.\par
This, second, direction of research is, first of all, of considerable theoretical interest, since the conditions for the realization of the hydrodynamic regime in a degenerate Fermi gas differ from those in the case of an ordinary, non-degenerate gas and liquid. The effects caused by external fields are often more pronounced in the ballistic regime than in the case of the local equilibrium hydrodynamic regime.
Finally, in most experiments in small magnetic fields, it is precisely the ballistic transport regime that is realized, since the mean free path relative to interparticle collisions turns out to be on the order of, or even larger than the characteristic spatial scales of the flow. As the magnetic field increases,
the cyclotron radius begins to play the role of the path length, and the conditions for the
applicability of the hydrodynamic description are satisfied.\par
In papers [3] and [11], the Knudsen regime of current flow in a long narrow two-dimensional sample with
diffusely scattering boundaries was considered, where the mean free path relative to
interparticle collisions is much larger than the sample width. The electron gas was considered
to be degenerate. The limit of arbitrarily weak electric and magnetic fields was studied, and the
magnetoresistance and Hall coefficient $R_H$ were found. It turned out, in particular, that $R_H$ in this
limit is half the value usual for Ohmic transport.\par
Bearing in mind that, in experiments, the properties of the edges of the sample may differ, in this work
we studied magnetotransport in ballistic regime in an
asymmetric sample, one of the edges of which is smooth, i.e. reflects electrons specularly,
while the other scatters them diffusely. It is shown that in this case the Hall coefficient is
anomalously large as compared to its standard value. Finally, at a semi-quantitative level, the situation is
discussed when all types of scattering can be neglected except for scattering at the edges of the
sample. Note that the anomalously large value of the Hall coefficient in the structure studied by
us is, apparently, among the so-called ballistic anomalies in magnetotransport, discussed in the
scientific literature in 1980 - 1990s (see, for example, a review [25]).

\section{Problem statement and basic equations}\par
We will study the electrical transport of a degenerate two-dimensional electron gas in a long narrow
sample with a width $W$ to which a time-independent uniform longitudinal electric
field $\textbf E_0$ and a magnetic field $\textbf B$ perpendicular to the sample plane are applied (see Fig.~1). The sample will
be assumed to be sufficiently clean, and the temperature sufficiently low, so that the electron-phonon
scattering can be neglected, and the mean free path relative to the scattering of
electrons by each other and by impurities is much larger than its width. One edge of the sample
is considered smooth, reflecting electrons "specularly", while the other is rough, scattering
them diffusely. We direct the axis $x$ along the field $\textbf E_0$ and align it with the smooth edge of the
sample, direct the axis $y$ into the sample, and the magnetic field along the axis $z$ (see Fig.~1). Bearing in
mind to calculate the linear response of the system to electric and magnetic fields, we will
consider them arbitrarily weak.\par We write the Boltzmann kinetic equation for the one-particle distribution function $f (\textbf r,\textbf v)$
in the form
\begin{equation}
\textbf v \frac{ \partial f}{ \partial \textbf r}- e \textbf E \frac{ \partial   f}{ \partial    \textbf p}
 + \omega_c \frac{ \partial    f }{ \partial     \varphi }= St_{ee}[f]+St_{imp}[f],
\end{equation}
where $\textbf v$ and $\textbf p=m\textbf v$ are the speed and momentum of the electron, $e>0$ is the magnitude of
its charge, $\textbf E$ is the electric field equal to the sum of longitudinal and Hall $\textbf E_H$  fields, $\varphi$ is  the
angle of the velocity vector measured from the ordinate axis,
$\omega_c=eB/mc$ is the cyclotron frequency, $St_{imp}[f] = -(f-f^0)/\tau_{imp}$ is the integral of collisions
with impurities, which we will assume to be short-range, $f^0$ is the symmetric part of the
distribution function, and  $St_{ee}[f]$ is the integral of electron-electron collisions, as which we will
use the model collision integral (see, for example, [26] and [3]), which in the simplest way takes into account the
conservation of the number of particles, energy and momentum in electron-electron collisions:
\begin{equation}
St_{ee}[f]=-\frac{1}{\tau_{ee}}(f-P_{01}[f]),
\end{equation}
where $P_{01}$ is the operator of projecting the function of the angular variable $\varphi$ onto the zero
and first harmonics.

Representing the distribution function in the form
\begin{equation}
f = f_F+ \frac{\partial     f_F}{\partial      \varepsilon}g,
\end{equation}
where $f_F$ is the equilibrium Fermi function, $\varepsilon$ is the electron energy, taking into account the
smallness of the perturbation, the degeneracy of the electron gas, and the independence of the
distribution function from $x$ due to the homogeneity of the system along the ordinate axis,
from (1) we obtain:
\begin{equation}
\begin{split}
\sin\varphi \frac{\partial   g}{ \partial y}- eE_0\cos\varphi- eE_H\sin\varphi+\frac{1}{R_c} \frac{\partial g}{\partial \varphi}\\=-\gamma_{ee}(g-g^0-g^s-g^c)-\gamma_{imp}(g-g^0),
\end{split}
\end{equation}
where $R_c = v_F/\omega _c$ is the cyclotron radius, $\gamma _ {ee} =1/l_{ee},~\gamma _ {imp} =1/l_{imp},~l_{ee/imp}=v_F\tau_{ee/imp}$,  $~g^0$ is the
symmetric part of the function $g$, $g^c$ is the projection $g$ onto the cosine, and $g^s$ - onto the
sine.
\begin{eqnarray}
 g^c(y,\varphi) =\frac{\cos\varphi}{\pi}\int\limits^{2\pi}_0f(y,\varphi') \cos\varphi' d \varphi',\nonumber\\
 g^s(y,\varphi) =\frac{\sin\varphi}{\pi}\int\limits^{2\pi}_0f(y,\varphi') \sin\varphi' d \varphi',\nonumber
\end{eqnarray}\par
The function $g^0(y)$ is responsible for the change in the concentration of electrons at a
given point, and through the functions $g^c$ and $g^s$ the densities of the longitudinal and
transverse currents are expressed, respectively. The boundary conditions for the function are
written as:
\begin{eqnarray}
f^+(0,\varphi)=f^-(0,-\varphi),~ 0 \leq \varphi \leq\pi,\\
f^-(W) =C^- =\frac{1}{2}\int\limits^{\pi}_0f^+(W)\sin\varphi d\varphi,
\end{eqnarray}
which means specular reflection from the bottom edge (Fig. 1) and diffuse - from the top [27]. Obviously, the function $g( y,\varphi )$ also satisfies similar
conditions. From (5) and (6) it can be seen, in particular, that the transverse current at the
edges of the sample is zero, and due to the continuity equation it is zero everywhere in the
sample, whence it follows that in the system under consideration $g^s=0$ .

 \begin{figure}[t!]
 \centerline{\includegraphics[width=1\linewidth]{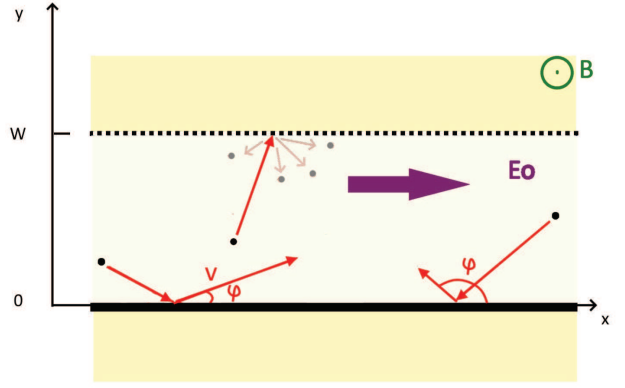}}
 \caption{
 Ballistic sample with a rough and a specular edges in external electric and magnetic fields.
 }
 \end{figure}

\section{Transport in the absence of a magnetic field}\par
In the absence of a magnetic field, the contribution to the symmetric part of the function
$g( y,\varphi )$ , which is linear in perturbation, is also equal to zero, i.e.$g^0=0$ . This follows from the
fact that for $B=0$, the distribution of electrons across the sample cannot depend on the
direction of the electric field $E_0$ applied to the sample (this is no longer the case $B \neq 0$ because
of the appearance of the Lorentz force). Therefore, at $B=0$ equation (4) takes the form
\begin{equation}
\sin\varphi \frac{ \partial  g_0}{ \partial  y}-eE_0\cos\varphi=-\gamma_{ee}(g_0-g_0^c)-\gamma_{imp}g_0.
\end{equation}
Finally, we will simplify it even further by omitting the function $g^0_c$ , which will be justified
below.\par
The solution of the resulting equation that satisfies boundary conditions (5) and (6) has the
form
\begin{eqnarray}
\begin{split}
g^{\pm}_0 (y,\varphi) =\frac{eE_0\cos\varphi}{\gamma}\Bigg[1-\exp\Bigg(-\gamma\frac{y\pm W}{\sin\varphi}\Bigg)\Bigg],\\ \gamma  =\gamma_{ee}+\gamma_{imp}.
\end{split}
\end{eqnarray}
Note that $g^-(W,\varphi)=0$. For the function $g^c_0(y,\varphi)=\frac{1}{\pi}\int\limits^{2\pi}_0 g_0(y,\varphi)\cos\varphi d\varphi$ from this
expression we obtain
\begin{equation}
g^c_0(\varphi)\approx \frac{4eE_0W}{\pi}\ln\Big(\frac{1}{\gamma W}\Big)\cos\varphi.
\end{equation}
Bearing in mind the inequality $\gamma |y\pm W|\ll 1$, from (8) and (9) it is easy to see what $g^c_0$ is greater
$g_0$ for all $\varphi$, except for narrow regions around the directions $\varphi = 0$ and $\varphi  = \pi$, where, on the
contrary, $g_0\gg g_0^c$ . In this regard, it may seem that the rejection $g_0^c$ in (7) was unjustified,
however, firstly, it is these narrow regions that made the main contribution to (9) and,
secondly, outside these regions, the entire right-hand side of (7) is small in parameter $\gamma W$ and
can be omitted. The correction $h( y,\varphi)$ to function (8), caused by taking into account $g_0^c$ in
equation (7), can be found by the perturbation method, writing $g( y,\varphi )$ in the form $g=g_0+h$
and substituting in (7) as $g_0^c$ expression (9). From the resulting equation, we find,
\begin{eqnarray}
h^{\pm}_0 (y,\varphi) =\frac{4eE_0W\cos\varphi}{\pi}\Bigg[1-\exp\Bigg(-\gamma\frac{y\pm W}{\sin\varphi}\Bigg)\Bigg]\nonumber,
\end{eqnarray}
which, due to the inequality $\gamma W\ll 1$ is small compared to (8). For current \[I =\frac{en_0}{\pi p_F} \int\limits^{W}_{0}dy\int\limits^{2\pi}_{0}g_0(y,\varphi)\cos\varphi d \varphi\]
and resistivity $\rho$ , from (8) we obtain
\begin{eqnarray}
I\approx \frac{4e^2n_0E_0W^2}{\pi p_F}\ln\Big(\frac{1}{\gamma W}\Big),~~ \rho \approx  \frac{\pi p_F}{4e^2n_0W\ln\Big(\frac{1}{\gamma W}\Big)}.
\end{eqnarray}
In the expression for the current density, we neglected the terms that depend on $y$ and do
not contain a large logarithm $\ln[1/(\gamma W)]$. Note that expression (10) for $\rho$ is half that obtained in
[3], which is not surprising, since there was considered a problem with two diffusely reflecting
edges.\par
Result (10) has a simple physical meaning. In the system under consideration, the electron
gas momentum can relax either upon collisions of electrons with a diffusely scattering edge of
the sample, or upon their collisions with impurities. Due to the condition $\gamma W \ll 1$ we have
adopted, typical electrons move along broken paths, randomly changing their direction of
motion after each collision with a diffusely scattering edge of the sample. The characteristic
momentum relaxation time of such electrons is of the order of $W/v_F \ll \tau_{ee,imp}$ and their
contribution to the conductivity is relatively small. The main contribution to the conductivity is
made by electrons moving at small angles to the axis $x$ .\par
 If $l_{imp}\ll l_{ee}$, then the relaxation length
of the momentum of such electrons is of the order of $l_{imp}$ or less, and they make a proportional
contribution \[\int\limits^1_{\gamma _ {imp}W}d\varphi/\varphi \sim \ln\Bigg(\frac{1}{\gamma_{imp} W}\Bigg)\] to the conductivity. In the opposite limiting case\linebreak
$l_{ee}\ll l_{imp}$ , an electron moving at a small angle, having passed a length of the order of $l_{ee}$ or
less, is scattered at a rough edge or collides with another electron, after which it becomes
typical and after a short time of the order $W/v_f$ is diffusely scattered at the rough edge of the
sample.
The corresponding contribution to the conductivity is proportional $\ln[1/(\gamma_{ee} W)]$.\par
 In the general
case, formula (10) is obtained. It is also clear from the last reasoning why, at $l_{imp}\gg l_{ee}$, the
outflow processes described by the integral of interparticle collisions $St_{ee} [ f ]$ play the main role
in the formation of the current, while the role of the incoming processes associated with the
function $g^c_0(y,\varphi$)is  small.\par
In sufficiently narrow samples, a different situation is possible. Due to the uncertainty
principle, the minimum transverse momentum of electrons is of the order $\hbar/W$, so that the
maximum time of motion without scattering before collision with the edge is of the order
$mW^2/\hbar$, and the minimum angle between the electron velocity vector and the ordinate axis is
of the order of the diffraction angle $\varphi \sim \hbar/Wp_F\sim \lambda_F/W $. As a result, if the inequality is satisfied
$1/k_F W\gg \gamma W$, i.e.$ W\ll\sqrt{l\lambda_F}$, where $l = 1/\gamma$, we again get formulas (10), which will enter
$\ln{(k_F W)}$ instead $\ln{[1/(\gamma W)]}$.
\section{Hall effect}\par
In this section of the article, we will find the Hall coefficient for our system in an arbitrarily
weak magnetic field. For this, it is necessary to take into account in the kinetic equation (4) the
terms with the Lorentz force and the Hall electric field. The function $g^0(y)$ is now nonzero,
since the action of the Lorentz force leads to a redistribution of electrons across the sample. Let
us write it in the form $g_0+g_1$ , where $g_1$ is the correction caused by the magnetic field,
which will be considered arbitrarily small and taken into account as a disturbance.We are
interested in the contribution to $g_1$ , linear in the magnetic field; therefore, in the term
$R_c^{-1} \partial{g}/\partial{\varphi}$ in equation (4), we can use function (8) as a function $g$ . In addition, since the
influence of the magnetic field on the current appears only in the second order in $\textbf B$ , we will
not take it into account by setting $g_1^c = 0$. Then from (4) we obtain the equation
\begin{equation}
\begin{array}{c}
\displaystyle
\sin\varphi \frac{\partial  g_1}{\partial  y} -\sin\varphi eE_H+\gamma(g_1-g^0_1)
%\\
%\\
%\displaystyle
=-\frac{1}{R_c} \frac{\partial g_1}{\partial  \varphi},\nonumber
\end{array}
\end{equation}
or
\begin{equation}
\begin{array}{c}
\displaystyle
\sin\varphi \frac{\partial  \tilde g_1}{ \partial  y} +\gamma(\tilde g_1-\tilde g^0_1)
%\\
%\\
%\displaystyle
=-\frac{1}{R_c} \frac{ \partial  \tilde g_1}{\partial  \varphi},
\end{array}
\end{equation}
where the function $\tilde g_1 =g_1+ e\Phi$ is introduced, $\Phi$ is the potential of the electric field.\par
The procedure for solving this equation is completely similar to the procedure for solving
equation (7): having omitted the function $\tilde g_1(y)$ in (11), we obtain the following expressions:
\begin{eqnarray}
\tilde g_1^+  \approx \frac{eE_0}{R_c}\Bigg\{ \frac{\sin\varphi}{\gamma^2}  -\exp\Bigg( -\gamma\frac{y+W}{\sin\varphi} \Bigg)\nonumber\\ \times\Bigg[   \frac{\sin\varphi}{\gamma^2}\Bigg(2\exp\Bigg(\frac{\gamma W}{\sin\varphi}\Bigg)-1\Bigg) + \frac{y-W}{\gamma}\nonumber\\-\frac{\cos^2\varphi}{2\sin^3\varphi}\big(y^2+2yW-W^2\big)-\frac{1}{2 \pi\gamma^2}\Bigg]\Bigg\}\nonumber\end{eqnarray}
and
\begin{eqnarray}
\tilde g_1^-  \approx \frac{eE_0}{R_c}\Bigg\{ \frac{\sin\varphi}{\gamma^2}  -\exp\Bigg( -\gamma\frac{y-W}{\sin\varphi} \Bigg)\nonumber\\ \times\Bigg[   \frac{\sin\varphi}{\gamma^2} + \frac{y-W}{\gamma}- \nonumber\\
-\frac{\cos^2\varphi}{2\sin^3\varphi}\big(y-W\big)^2-\frac{1}{2\pi\gamma^2}\Bigg]\Bigg\}.
\end{eqnarray}
Then we show that the correction arising from the account $\tilde g_1(y)$ is parametrically small
(see Appendix A).
Substituting these expressions in
\[
2 \pi \tilde g^0_1 (y)  = \int ^{2\pi}_0 g_1(y,\varphi) d\varphi
\]
 and keeping only the main
contribution, we find
\begin{equation}
\tilde g_1^0\approx  -\frac{eE_0}{\pi R_c}\Bigg[\frac{W^2}{\gamma^2(y+W)^2}-\frac{1}{2\gamma^2}\Bigg].
\end{equation}
The function $\tilde g_1^0/e $ is the electrochemical potential $\Psi(y)$, its minus derivative is equal to
the Hall field, and the difference in values at the edges of the sample is measured with a
voltmeter and is equal to the Hall voltage, $U_H  = \Psi(W)-\Psi(0)$. Therefore, we have
\begin{eqnarray}
E_H\approx -\frac{2E_0 W^2}{\pi R_c\gamma (y+W)^3},~~ U_H\approx  -\frac{3E_0}{4\pi R_C\gamma^2}.
\end{eqnarray}
From here and from (10) for the Hall coefficient we obtain
\begin{eqnarray}
R_H = \frac{U_H}{BI}  = \frac{3}{16\gamma^2W^2\ln(1/\gamma W)}\frac{1}{en_0c}\gg\frac{1}{en_0c}.
\end{eqnarray}

Near the points $\varphi = 0$ and $\varphi = \pi$ functions (12) have singularities of the form $1/\varphi^3$ and
$1/(\varphi-\pi)^3$, making the main contribution (13) to the function $\tilde g^0_1(y)$, the divergences arising in
this case are "cut off" by exponential factors $\exp[-\gamma(y\pm W)/\sin\varphi]$. Note that in the case of a
symmetric structure, which was studied in papers [3] and [11], the main contributions to $\tilde g^0_1(y)$ from functions
$g_1^+$ and $g_1^-$ cancel each other, and the Hall coefficient turns out to be equal $1/2en_0c$ .
In narrow samples, $W\ll \sqrt{l\lambda_F}$, the divergences are "cut off" due to the principle of
uncertainty at angles $|\varphi|$ and $|\varphi-\pi|$ order $1/Wk_F$ , and as a result, the following expression is
obtained for the Hall coefficient
\begin{equation}
R_H = A\frac{(k_FW)^2}{\ln(k_FW)}\cdot\frac{1}{en_0c},
\end{equation}
where $A$ is a numerical coefficient of the order of unity, which remains undefined within the
framework of our consideration.\par

In conclusion of this section, we will explain the anomalously large value of the Hall
coefficient in the asymmetric structure we studied. To do this, let's approach the problem from
a slightly different point of view. We represent the distribution function in the form
$f(\varepsilon,\textbf r,\varphi) = f_0(\varepsilon,\textbf r)+\tilde f(\varepsilon,\textbf r, \varphi), \int_0^{2\pi}\tilde fd\varphi = 0$, multiply equation (1) by $p_y$ and integrate over
$2d^2\textbf p/(2\pi\hbar)^2$. It will turn out
\begin{equation}
\begin{split}
\frac{ \partial  \overline\varepsilon }{\partial  y}+enE_y+\frac{ \partial  \Pi_{yx} }{ \partial  x}+\frac{ \partial  \Pi_{yy} }{ \partial y}-m\omega_cnV_x+\frac{mn}{\tau_{imp}}V_y,
\end{split}
\end{equation}
where the notation
\begin{equation}
\begin{split}
n(\bm r)=\int f_0\frac{2d^2\bm p}{(2\pi\hbar)^2}, ~\overline\varepsilon = \int f_0\varepsilon \frac{2d^2\bm p}{(2\pi\hbar)^2},\\ \bm V(\bm r) =\frac{1}{n(\bm r)} \int \tilde f\bm v \frac{2d^2\bm p}{(2\pi\hbar)^2},\\~ \Pi_{ik} \equiv m\int \tilde f v_i v_k \frac{2d^2\bm p}{(2\pi\hbar)^2}
\end{split}
\end{equation}
is introduced. The equation
\begin{equation}
\begin{split}
\frac{ \partial  \overline\varepsilon }{ \partial  x}+enE_x+\frac{ \partial  \Pi_{xx} }{ \partial  x}+\frac{ \partial  \Pi_{xy} }{ \partial y}+m\omega_cnV_y+\frac{mn}{\tau_{imp}}V_x
\end{split}
\end{equation}
is obtained similarly. Equations (17) and (19) are exact. It is convenient to write them in the
form
\begin{eqnarray}
enE_x+\frac{ \partial  \overline\varepsilon }{\partial  x} + \nabla\cdot \boldsymbol\Pi_x +m \omega_c nV_y +\frac{mn}{\tau_{imp}}V_x =  0 ,\nonumber\\
enE_y+\frac{ \partial  \overline\varepsilon }{ \partial  y} +\nabla\cdot \boldsymbol\Pi_y -m \omega_c nV_x +\frac{mn}{\tau_{imp}}V_y =  0 ,\nonumber\\
\boldsymbol\Pi_i = \int \tilde f v_i\bm v\frac{2d^2\bm p}{(2\pi\hbar)^2}.
\end{eqnarray}
In equilibrium, the electric field and average velocity $\bm V(\bm r)$
are equal to zero; therefore, in the
linear approximation, these equations take the form
\begin{eqnarray}
en_0E_x+\frac{ \partial  \overline\varepsilon }{ \partial  x} +\nabla\cdot\boldsymbol\Pi_x +m \omega_c n_0V_y +\frac{mn_0}{\tau_{imp}}V_x =  0 ,\nonumber\\
en_0E_y+\frac{ \partial  \overline\varepsilon }{\partial  y} +\nabla\cdot\boldsymbol\Pi_y -m \omega_c n_0V_x +\frac{mn_0}{\tau_{imp}}V_y =  0 .
\end{eqnarray}
Let us introduce the electric $\Phi$ and electrochemical potentials $\Psi  = \Phi - \overline\varepsilon/en_0$ and rewrite the
equations (21) through them
\begin{eqnarray}
en_0\frac{ \partial  \Psi}{ \partial  x} -\frac{n_0eB}{c}V_y +\frac{mn_0}{\tau_{imp}}V_x=\nabla\cdot\boldsymbol\Pi_x   ,\nonumber\\
en_0\frac{ \partial  \Psi}{ \partial  y} +\frac{n_0eB}{c}V_x +\frac{mn_0}{\tau_{imp}}V_y=\nabla\cdot\boldsymbol\Pi_y   .
\end{eqnarray}
The function $\Psi  = \Phi - \overline\varepsilon/en_0$ is the same as the function $\Psi  = \tilde g^0_1/e = \Phi +g^0_1/e$
we introduced above. Really:
\begin{equation}
\begin{split}
\overline\varepsilon \equiv \int f_0 \nu \varepsilon d\varepsilon=\int \frac{ \partial  f_F}{ \partial  \varepsilon} g_1^0\nu \varepsilon d\varepsilon  = -\nu \varepsilon_F g_1^0 = -n_0g_1^0,
\end{split}
\end{equation}
where $\nu = m/\pi\hbar^2$ is the density of states. \par
In the left-hand sides of equations (22), there are forces acting on a unit volume of the
electron gas in $x$ and $y$ out of directions, and on the right, divergence of the momentum flow in
the same directions. We are interested in the second equation. Let us rewrite it, taking into
account that in our problem there is no dependence on the coordinate $x$ and the average
velocity along the axis $y$ is zero
\begin{equation}
en_0\frac{ \partial  \Psi}{ \partial  y} +\frac{n_0eB}{c}V_x =\frac{ \partial  \Pi_{yy}}{ \partial  y}   ,\nonumber
\end{equation}
and integrate across the sample. It turns out:
\begin{equation}
U_H = \frac{1}{en_0}\big[ \Pi_{yy}(W)-\Pi_{yy}(0)\big]+\frac{BI}{en_0c}.
\end{equation}
Substituting now the function (12) multiplied $\partial f_f/\partial\varepsilon$ into the definition $\Pi_{yy}$ and performing
the integration, we again obtain expression (14) for the Hall voltage (note that the contribution
to the integral of the symmetric part of the function $\tilde g_1$
is zero). It can be seen from the
foregoing that the Hall voltage arising in our asymmetric sample is primarily intended to
compensate for the gradient of the transverse-pulse flux density across the sample, and not the
Lorentz force, as is usually the case in systems with an ohmic current flow. A nonzero gradient
of the momentum flux density in the transverse direction also arises in a symmetric structure,
but it is of the same order of magnitude as the Lorentz force and leads to a halving of the Hall
coefficient compared to its standard value.
\section{Conclusion}
In this work, we studied the Knudsen regime of a degenerate electron gas flow in a Hall sample,
one edge of which reflects electrons diffusely, and the other specularly. The collisions of
electrons with each other and their collisions with impurities were taken into account.
It is shown that, in such an asymmetric sample, the Hall coefficient is parametrically large in
comparison with its standard value $R^0_H =1/en_0c$ :
$R_H/R^0_H \sim l^2/W^2\ln(l/W) $ , where $l$ is the
effective momentum relaxation length $l \gg W$. In
addition, the situation is discussed at a semiquantitative level when all types of scattering can
be neglected except for scattering at the edges of the sample.\par
Finally, we note that, when deriving formula (15) for the Hall resistance, the role of ``skipping orbits''
near the rough edge was not thoroughly analyzed, which can lead to a change in the numerical factor in
the expression for the Hall resistance [28].
\begin{acknowledgments}
	We thank P. S. Alekseev for valuable discusions. Yu.A. also thanks to his teachers M. E. Kompan, N. M. Khimin, M. G. Ivanov. This work was supported by the Russian Science Foundation (grant No.~17-12-01182-c).
\end{acknowledgments}

\section{Appendix A}
Having performed the calculations described in the text, we obtain for corrections to functions (12)
\begin{equation}
\begin{split}
h_1^+(y,\varphi) = \frac{\gamma}{\sin\varphi}\Bigg[\int\limits^y_0 \exp\Bigg(-\gamma\frac{y-y'}{\sin\varphi}\Bigg)\tilde g_1^0(y')dy'\\+\int\limits^W_0 \exp\Bigg(-\gamma\frac{y+y'}{\sin\varphi}\Bigg)\tilde g_1^0(y')dy'\Bigg],\\
h_1^+(y,\varphi) = \frac{\gamma}{\sin\varphi}\int\limits^y_W \exp\Bigg(-\gamma\frac{y-y'}{\sin\varphi}\Bigg)\tilde g_1^0dy',
\end{split}
\end{equation}
where $\tilde g^0_1 =\tilde g^0_1(y') $
is the symmetric part of function (12). Having integrated it over and over, we obtain for the correction to the Hall coefficient
\begin{equation}
\delta R_H \approx -\frac{1}{8\pi\gamma W}\cdot\frac{1}{en_0c},\nonumber
\end{equation}
which is parametrically less than expression (15).


\begin{thebibliography}{}
\bibitem{27}
	Note that the condition of equality of the concentrations of particles flying up to a rough boundary and reflected from it, encountered in the literature, is erroneous; in particular, it leads to a nonzero particle flux through this boundary.
\end{thebibliography}
\end{document}